\documentclass[10pt,twocolumn,twoside]{article}
\usepackage{authblk}
\usepackage[numbers]{natbib}
\usepackage[T1]{fontenc}

\usepackage{url,lineno,microtype} 
\usepackage[version=4]{mhchem}
\usepackage{multirow}
\usepackage{siunitx}
\usepackage{enumitem}
\usepackage{textcomp,mathcomp,bbold,times,graphicx}
\usepackage{amssymb,tikz}
\usepackage{widetext}
\usepackage[hyperindex=true]{hyperref}
\usepackage{color}
\definecolor{linkcol}{rgb}{0,0,0.4}
\definecolor{citecol}{RGB}{0,20,95}
\definecolor{seagreen}{rgb}{0.18, 0.55, 0.34}
\definecolor{darkred}{rgb}{0.7, 0.11, 0.11}
\hypersetup
{
bookmarksopen=true,
pdftitle="Manuscript title",
pdfauthor="Your name",
pdfsubject="Manuscript topic in a few words", 
pdfmenubar=true, 
pdfhighlight=/O, 
colorlinks=true, 
pdfpagemode=UseNone, 
pdfpagelayout=SinglePage, 
pdffitwindow=true, 
linkcolor=linkcol, 
citecolor=citecol, 
urlcolor=linkcol 
}

\graphicspath{{PNG}}

\newcommand{\OmegaD}[1]{\ensuremath{{\Omega^{#1 D}}}}

\RequirePackage{graphicx,xcolor}
\RequirePackage[english,linesnumbered,algoruled]{algorithm2e}
\RequirePackage{colortbl}
\RequirePackage{booktabs}
\RequirePackage[noend]{algpseudocode}
\RequirePackage{changepage}
\RequirePackage[twoside,%
				letterpaper,includeheadfoot,%
				layoutsize={8.125in,10.875in},%
                layouthoffset=0.1875in,%
                layoutvoffset=0.0625in,%
                left=38.5pt,%
                right=43pt,%
                top=43pt,
                bottom=44pt,%
                headheight=0pt,
                headsep=-2pt,%
                footskip=25pt,
                marginparwidth=38pt]{geometry}
\RequirePackage[labelfont={bf},%
                labelsep=period,%
                figurename=Figure]{caption}
\begin{document}

\title{\bf When Cubic Law and Darcy Fail: Bayesian Correction of Model Misspecification in Fracture Conductivities}
\author[a]{Sarah Perez\footnote{Corresponding author: S.Perez@hw.ac.uk}$^,$}
\author[b]{Florian Doster}
\author[b]{Julien Maes}
\author[b]{Hannah Menke}
\author[b]{Ahmed ElSheikh}
\author[a]{Andreas Busch}
\affil[a]{The Lyell Centre, Heriot-Watt University, Edinburgh, UK}
\affil[b]{Institute of GeoEnergy Engineering, Heriot-Watt University, Edinburgh, UK}
\maketitle

\begin{abstract}
Structural uncertainties and unresolved features in fault zones hinder the assessment of leakage risks in subsurface \ce{CO2} storage. Understanding multi-scale uncertainties in fracture network conductivity is crucial for mitigating risks and reliably modelling upscaled fault leakage rates. Conventional models, such as the Cubic Law, which is based on mechanical aperture measurements, often neglect fracture roughness, leading to model misspecifications and inaccurate conductivity estimates. Here, we develop a physics-informed, AI-driven correction of these model misspecifications by automatically integrating roughness effects and small-scale structural uncertainties. Using Bayesian inference combined with data-driven and geometric corrections, we reconstruct local hydraulic aperture fields that reliably estimate fracture conductivities. By leveraging interactions across scales, we improve upon traditional empirical corrections and provide a framework for propagating uncertainties from individual fractures to network scales. Our approach thereby supports robust calibration of conductivity ranges for fault leakage sensitivity analyses, offering a scalable solution for subsurface risk assessment.\\

\end{abstract}
\vspace{-5mm}
\section*{Significance Statement}

Safely storing \ce{CO2} in the subsurface is crucial for mitigating climate change, but predicting potential leakage through fault zones remains challenging due to uncertainties in subsurface structures. Fault zones are complex and span a wide range of scales, with small-scale features significantly affecting fluid flow. However, these small structures are often ignored in current models. This study develops an AI-powered framework to address these uncertainties, combining data-driven corrections with physics-based insights to estimate fracture conductivity more accurately. By capturing the effects of fracture roughness and propagating their uncertainties to network scales, the approach aims to improve predictions of leakage risks. This work will help bridge the gap between small-scale fracture behaviours and large-scale fault models, enhancing the reliability of \ce{CO2} storage simulations and contributing to safer, more sustainable climate solutions.

\section{Introduction}

Carbon Capture and Storage (CCS) is a mature technology supporting the decarbonisation of energy-intensive industries and mitigating climate change \cite{BACHU2015, PNAS_2023}. Despite low-permeability seals, leakage along wells and through the geological overburden remains a low, yet unquantified risk. Fault zones can form highly conductive pathways that compromise seal integrity and increase leakage risks \cite{RIZZO2024, BISDOM2024}. Understanding the hydraulic behaviour of fault zones is critical for de-risking CCS operations and quantifying fault-related leakage rates.

Fault zones are geologically complex, typically comprising a low-permeability fault core surrounded by a damage zone consisting of fracture networks \cite{RIZZO2024}. The conductivity of this damage zone is governed by a complex interplay of fracture permeability and connectivity, compounded by multi-scale interactions with the surrounding matrix that result in a broad range of fluid migration rates \cite{Viswanathan_2022, PNAS_2024_Frac_Net}. Accurately characterising these interactions is challenging due to structural uncertainties, particularly geological features below the resolution of seismic imaging, which can significantly complicate leakage modelling \cite{GONG_2019, FANG2022, DASHTI2024, Xiaopeng_2024}. Sensitivity analyses are widely used to assess the impact of these uncertainties on leakage rates. However, the parametrisation of geological distributions, including fracture network conductivity, is often manually tuned by reservoir engineers, highlighting the need for reliable calibration of the models \cite{Jayne2019, SUN2023}.

Smaller-scale uncertainties are often overlooked when calibrating probabilistic distributions for large-scale sensitivity analyses, and their effects remain poorly understood. Several key questions remain largely unexplored: Can we achieve reliable calibration of fault conductivity distributions through a meaningful analysis of uncertainties across different scales? Can we develop a framework that integrates both structural and modelling uncertainties at the individual fracture scale and effectively propagates them to fault leakage rate distributions? We address this gap by establishing the foundation for a robust upscaling strategy to assess uncertainties in fault-related leakage.

We propose leveraging the interplay between individual fractures, fracture networks, and fault zones to understand uncertainty propagation across scales. Conductivity is particularly sensitive to local geometric features, such as aperture distribution and surface roughness \cite{Tomos_2021, HE2021_ML}. However, conventional models often assume smooth, parallel fracture walls and neglect small-scale uncertainties like roughness. For example, the Cubic Law that relies on the average  structural width of a fracture, overlooks key differences between mechanical and hydraulic apertures, the latter being the effective width contributing to fluid flow. Although empirical corrections to the Cubic Law have been proposed, they remain highly parametrised and difficult to apply to realistic fracture networks \cite{Rasouli2011, Xie2015, HE2021}. These modelling misspecifications, stemming from oversimplified assumptions or measurement limitations, can bias conductivity estimates and undermine their reliability at the fracture and fracture network scales \cite{FANG2022}. It is therefore crucial to investigate the uncertainties arising from these misspecifications across scales to support robust decision-making regarding fault-related leakage \cite{FANG2022}.

To address these challenges, we present a probabilistic approach that combines physics-based and data-driven methods, leveraging the machine learning framework developed by Perez et al. \cite{Perez2023}. We introduce an automatic and geometric correction of model misspecifications to infer the latent (unobservable) hydraulic aperture and its uncertainties. Since the hydraulic aperture is a flow-related concept that cannot be directly measured from fracture geometries, we develop a proxy that reconstructs a local hydraulic field, and its associated permeability field, by automatically accounting for roughness effects. To our knowledge, the novelty of this work lies in explicitly modelling hydraulic apertures as spatially varying latent fields, inferred from flow physics and geometric data. While previous studies have estimated scalar hydraulic apertures or applied local corrections to permeability, none have reconstructed hydraulic aperture as a probabilistic spatial field with physically grounded constraints and uncertainty quantification. At its core, our method relies on a multi-objective Bayesian inference formulation that integrates mechanical aperture data and physics-based constraints, regularising small-scale behaviour and uncertainties. This correction adaptively quantifies deviations between hydraulic and mechanical apertures, yielding local permeability fields that align with the effective permeability, representing the upscaled behaviour. We validate the approach on synthetic fractures with identical mean mechanical apertures but varying roughness, demonstrating improved estimates of hydraulic aperture and conductivity where both the Cubic Law and Darcy-based models fail. Finally, our local, geometry-aware correction benefits from enhanced generalisation over empirical methods, enabling reliable upscaling at the network scale. Our key contributions include:
\begin{itemize}
\item We demonstrate that models relying on mechanical aperture fields, whether applying the Cubic Law or Darcy upscaling, systematically overlook roughness effects and overestimate fracture conductivity, even in simplified synthetic geometries. This underscores the need to quantify and correct model misspecifications at the fracture scale to support reliable uncertainty propagation in complex fracture networks.
\item We introduce a data-driven probabilistic framework that combines physical priors with geometric insights to automatically correct fracture conductivity models, establishing a foundation for robust generalisation and upscaling to network scales.
\item Our approach produces local hydraulic aperture and corrected permeability maps that are consistent with effective upscaling. These maps can be integrated into fracture network models and directly used for Darcy upscaling, yielding more reliable network conductivity estimates with robust uncertainty bounds to enable meaningful sensitivity analyses of fault-zone leakage rates.
\end{itemize}

\section{Method}

\subsection{Evidence of Model Misspecifications in Fracture Conductivity}

To investigate model misspecifications in fracture conductivity, we consider a synthetic dataset designed to mimic realistic geological fractures while enabling systematic evaluation of roughness effects. While high-resolution micro-computed tomography scans ($\mu$CT) provide detailed descriptions of fracture networks, they usually span billions of pixels, rendering Direct Numerical Simulations (DNS) of flow computationally impractical, especially at large scales \cite{Landry_2012, Adenike_2015}. Modelling alternatives typically rely on mechanical aperture maps and simplified models, such as the Cubic Law \cite{Zimmerman1996}, despite their limitations in accounting for roughness and multi-scale effects \cite{Konzuk_2004, CRANDALL2010}. This introduces significant model misspecifications, which in turn compromise estimates of fracture conductivity and their associated uncertainties.

\begin{figure*}[ht!]
\centering
\includegraphics[width=\linewidth,trim={0.3cm 0 0.4cm 0.8cm},clip]{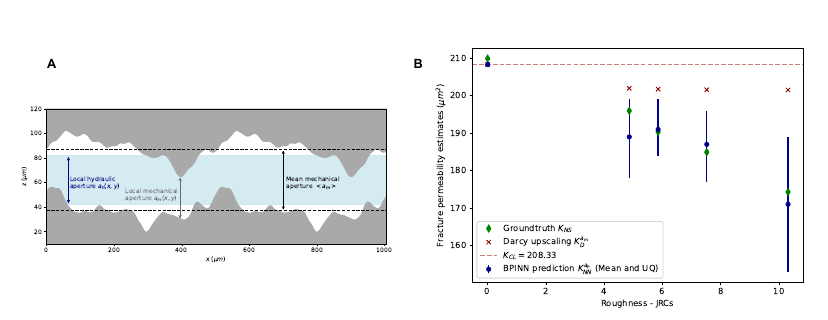}
\caption{(A) Schematic illustrating how the hydraulic aperture field $a_h(x,y)$ can differ from the geometrically measured mechanical aperture field $a_m(x,y)$ due to surface roughness. The two horizontal dashed lines represent the parallel planes of the fracture geometry for the Cubic Law approximation, with the mean mechanical aperture $<a_m>$. (B) Comparison of fracture permeabilities from different modelling assumptions for various roughness patterns identified by their JRCs, highlighting modelling misspecifications and correction. The models include direct Cubic Law approximation $K_{CL}$ (red dashed line); Darcy flow upscaling $K_{D}^{\,a_m}$, based on the 2D mechanical aperture map; the groundtruth $K_{NS}$, based on the 3D fracture geometry; and the Bayesian PINN correction, which provides an upscaled mean estimate (blue dots) with associated uncertainty ranges. Detailed values are provided in Table S1 in SI.}
\label{fig:apertures_models}
\end{figure*}

Our dataset comprises four fracture geometries, obtained from the Digital Rock Portal \cite{Frac_Dataset}, each with dimensions of $256\times128\times30$ units, isotropic resolution of $\delta x = 3.94\,\mu m$ and a common mechanical aperture $<a_m>$ of $50\,\mu m$. The notation $<a_m>$ refers to the arithmetic mean of the 2D mechanical aperture field $a_m(x,y)$, describing the fracture width in the $z$ direction for each point $(x,y)\in \OmegaD{2}$, with $\OmegaD{2}$ the 2D projection of the 3D fracture domain $\OmegaD{3}$. The arithmetic mean is thus defined as:
\begin{equation}
    <a_m> = \displaystyle \frac{1}{|\OmegaD{2}|}\int_\OmegaD{2} a_m(x,y)dx dy.
    \label{eq:mean_am}
\end{equation}
The fracture geometries exhibit varying wall roughness, characterised by their Joint Roughness Coefficients (JRCs) ranging from $4.86$ to $10.31$, with higher values indicating rougher surfaces \cite{MYERS1962, LI2015}. Further details on the dataset generation are provided in Guiltinan et al. \cite{Guiltinan_2021}. The fracture profiles are finally corrupted with Gaussian noise to mimic artefacts commonly encountered in $\mu$CT datasets \cite{Banhart_2008}.

We apply several modelling approximations on this dataset to assess their robustness to roughness effects, benchmarking them against the effective permeability $K_{NS}$. This reference permeability is computed via DNS by solving the Stokes equation within the 3D fracture geometries. By fully resolving the flow field, $K_{NS}$ remains the most reliable estimate of fracture permeability, and serves as the groundtruth for our comparative analysis. Details on the computation of $K_{NS}$ are provided in Text S1 in Supporting Information (SI), and we refer to Perez et al. \cite{Perez2022} for a full description of the DNS method.

First, the empirical Cubic Law estimates fracture permeabilities, denoted as $K_{CL}$, based on the arithmetic average mechanical aperture $<a_m>$ through the relation 
\begin{equation}
    K_{CL}=\frac{<a_m >^2}{12}.
\label{eq:Cubic_Law}
\end{equation}
This model assumes smooth and parallel fracture walls with no roughness, and raises questions about the validity of the Cubic Law  for realistic fracture geometries, especially considering the discrepancies that roughness can induce between the mechanical and hydraulic apertures \cite{Tomos_2021} (see Figure \ref{fig:apertures_models}). Furthermore, the Cubic Law implies that conductivity is independent of the flow direction, \emph{i.e.}, it assumes the fracture to be isotropic, a condition rarely met in natural fractures. According to the Cubic Law, the different fractures in the dataset present the same estimated permeability of $K_{CL}\simeq208\,\mu m^2$. However, flow simulations using DNS in $\OmegaD{3}$ reveal non-negligible deviations in the effective fracture permeability $K_{NS}$ across the different JRCs, with discrepancies ranging from $6\%$ to $16\%$ (see Table S1 and Figure S1 in SI). 

\begin{figure*}[!ht]
\centering
\includegraphics[width=\linewidth]{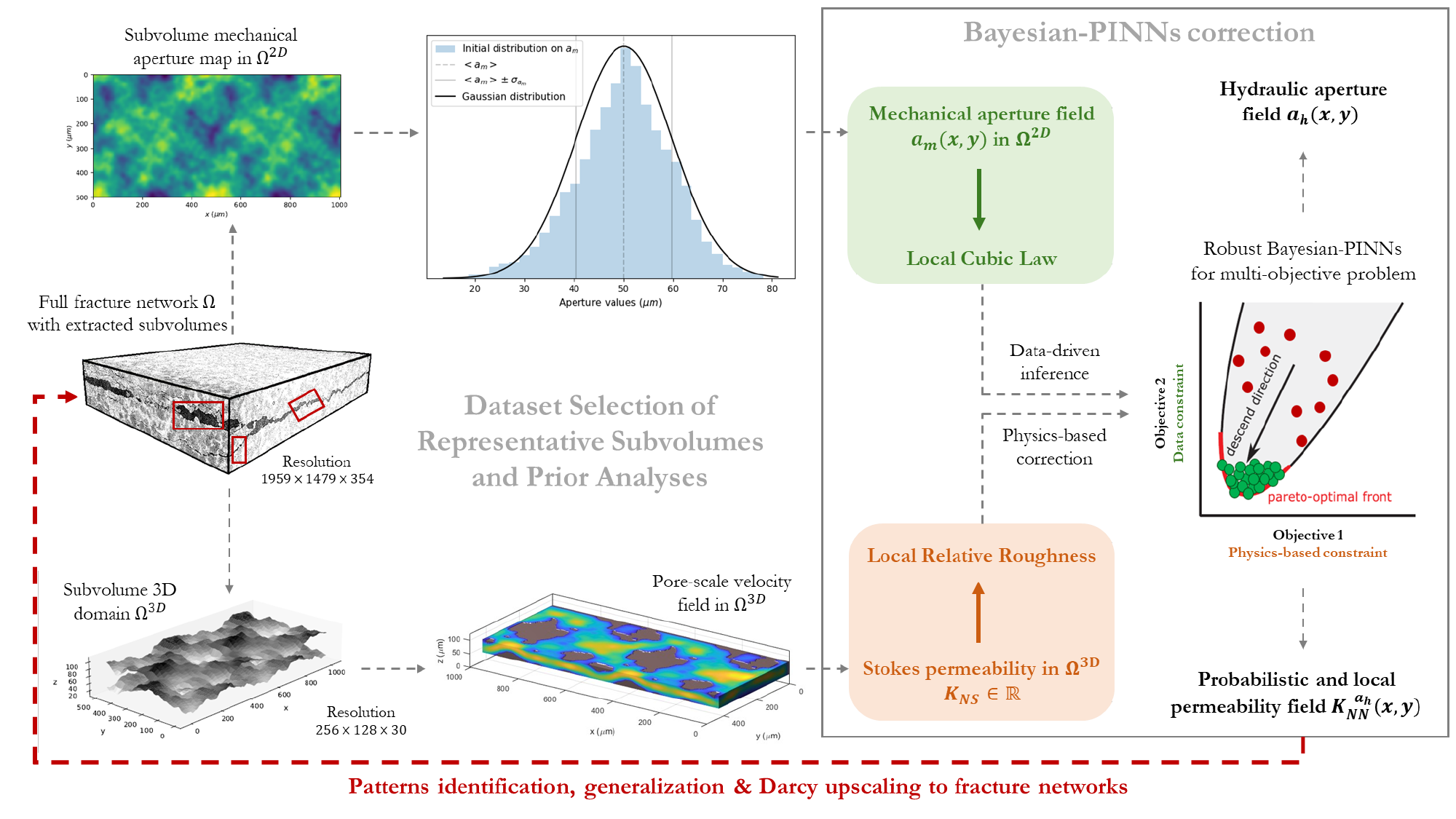}
\caption{Overview of the workflow for correcting model misspecification in estimating the conductivity of rough fractures and complex fracture networks. 
Starting from the 3D geometry of a fracture network obtained via $\mu$CT, we infer hydraulic aperture, $a_h(x,y)$, and local permeability, $K_{NN}^{\,a_h}(x,y)$, fields over a few extracted representative subvolumes. The Bayesian inference is achieved through our AI-driven uncertainty quantification methodology, combining data-driven and physics-based approaches. Efficient exploration of the Pareto front, which involves sampling the green points while avoiding unbalanced conditions (red points), is ensured through a robust B-PINN formulation for multi-objective inference. The data-driven part relies on input measurements of the mechanical aperture field and identifies limitations of the local Cubic Law. The physics-based correction benefits from the input upscaled permeability estimate $K_{NS}$, obtained by solving the Stokes equation on the subvolumes prior to the inference, to incorporate the effects of local relative roughness. Our method establishes a mapping between unreliable mechanical aperture data and corrected hydraulic aperture and permeability fields, including their associated uncertainties. This mapping can then be integrated with pattern recognition on realistic fracture networks together with Darcy upscaling, using the corrected permeability maps $K_{NN}^{\,a_h}(x,y)$, to provide uncertainty bounds for fracture network conductivity on a larger scale, without the need for additional DNS.}
\label{fig:workflow}
\end{figure*}

While the averaged behaviour of the Cubic Law represents a major misspecification, we also investigate a more refined modelling alternative that assumes local validity of the Cubic Law. The latter considers the local mechanical aperture $a_m(x,y)$ to derive a pointwise permeability map:
\begin{equation}
    K_{CL}^{\,a_m}(x,y) = \frac{a_m (x,y)^2}{12}.
\label{eq:Perm_map_CL}
\end{equation}
Fracture permeability can then be estimated by applying Darcy flow-based upscaling to this 2D permeability field, incorporating flow direction into the conductivity evaluation. The Darcy permeability values, denoted $K_D^{\,a_m}$, are computed using the MATLAB Reservoir Simulation Toolbox (MRST) for single-phase flow \cite{Lie_2019}. Modelling details are provided in Text S2 in SI. $K_D^{\,a_m}$ is finally compared with the Cubic Law approximation $K_{CL}$ and the effective permeability $K_{NS}$ for the different JRCs. The results highlight that the local Cubic Law model combined with Darcy upscaling tend to reduce overestimation on the fracture permeability, compared to $K_{CL}$. However, Figure \ref{fig:apertures_models}B reveals persistent deviations between $K_D^{\,a_m}$ and $K_{NS}$, ranging from approximately $3\%$ to $13.5\%$ across the different roughness levels. This modelling alternative also fails to capture the roughness effect: in Figure \ref{fig:apertures_models}B, the permeability estimates vary only slightly, remaining near $K_D^{\,a_m} \simeq 201\,\mu m^2$ despite increasing roughness. This highlights that the local application of the Cubic Law neither captures nor accounts for the geometrical characteristics of the fracture, which are discarded when the 3D domain is projected onto aperture maps. While Figure \ref{fig:apertures_models}B only shows relative deviations up to 16\%, further experiments on synthetic geometries reveal errors approaching 40\% for narrower fractures (see Figure S2 in SI). In real fractured sandstones, roughness effects can result in conductivity reductions of up to 35-fold \cite{CRANDALL2010} and even two orders of magnitude \cite{Adenike_2015}, underscoring the severity of model misspecification. Even small deviations can be consequential in fracture networks or fault systems, where multi-scale heterogeneities can amplify local conductivity errors, shift preferential flow pathways, and trigger large-scale uncertainties in predicted leakage rates.

Overall, we observe the following hierarchical relationships in the modelling of fracture permeability: $K_{CL} > K_D^{\,a_m} > K_{NS}$. These findings highlight the limitations of relying solely on mechanical aperture maps, and underscore the need for more advanced correction framework across the different scales.

\subsection{Automatic AI-driven Correction with Uncertainty Quantification}

We propose a probabilistic, AI-driven correction framework to address model misspecifications in fracture conductivity. Our focus is on reconstructing the latent hydraulic aperture field $a_h(x,y)$ and its uncertainties using a multi-objective Bayesian inference approach. In this context, the challenge lies in properly accounting for the imbalance in magnitude between different objectives --- such as data fidelity and physical consistency --- that must be fulfilled simultaneously.  Characterising this scale imbalance is crucial, as it impacts the convergence of the inference and robustness of the resulting uncertainty estimates.

To address this challenge, we build on recent advances in deep learning surrogates. Deep learning surrogate models based on Bayesian Physics-Informed Neural Networks (B-PINNs) have shown strong potential for uncertainty quantification \cite{YANG2021, LINKA2022, Perez2024}, and are increasingly used to accelerate Bayesian inference across a variety of scientific applications \cite{Molnar_2022, BPINN_subsurface_2023, Li_2024}. Our correction framework leverages a robust B-PINN formulation, enhanced with the Adaptively Weighted Hamiltonian Monte Carlo (AW-HMC) sampler introduced by Perez et al. \cite{Perez2023}. This approach automatically balances the influence of data and physical constraints by dynamically adjusting task weights during posterior sampling in the Bayesian process. It efficiently explores the set of optimal trade-off solutions, \emph{i.e.}, the Pareto front, where no objective can be improved without compromising another, as illustrated in Figure \ref{fig:workflow}. We capture the full distribution of solutions that accounts for the uncertainties of each specific objective, thereby highlighting the most uncertain tasks. As a result, we obtain a robust inference of the hydraulic aperture and local permeability fields, even under significant modelling uncertainties and imbalanced objectives.

We assume that synthetic or experimental $\mu$CT images of the fracture geometries are available in the 3D domain $\OmegaD{3}$, together with mechanical aperture maps defined on $\OmegaD{2}$. While this section focuses on the previously introduced synthetic fractures, larger and more realistic fracture networks are handled using an ensemble of representative subvolumes (represented by the red squares in Figure~\ref{fig:workflow}) designed to capture local patterns in aperture and roughness. These subvolumes, mapping local mechanical and hydraulic apertures via Bayesian inference, will subsequently serve as training data for efficient uncertainty upscaling, eliminating the need for DNS over the entire fracture network geometry. In this context, the input dataset $\mathcal{D}$ for the B-PINNs correction comprises $N_{obs} \simeq 10,000$ scattered and noisy mechanical aperture values $a_m(x_i, y_i)$ sampled from high-resolution grids ($256 \times 128$ points) for each synthetic fracture. Once selected, this training dataset remains fixed during inference, while final predictions are validated across the entire computational domain $\OmegaD{2}$. 

Unlike traditional approaches that treat hydraulic aperture as a scalar, we introduce a spatially resolved local hydraulic aperture field $a_h(x, y)$, formulated as a latent variable in a Bayesian inference problem. The Bayesian inference integrates three key components:
\vspace{-0.3cm}

\begin{enumerate}[label=(\roman*)]
    \item \textbf{Data fitting}\\
The local mechanical aperture, $a_m(x_i, y_i)$, is treated as noisy observations of the hydraulic aperture field:
\begin{equation}
    a_m(x_i,y_i) = a_h(x_i, y_i) + \xi_d(x_i, y_i),
\end{equation}
where $\xi_d\sim \mathcal{N}(0, \sigma_d^2I)$ and $\sigma_d$ is an unknown standard deviation, referring to data uncertainties. The noise term, $\xi_d$, captures both measurement errors and conceptual differences between the mechanical and hydraulic aperture fields. 

    \item \textbf{Physical bound}\\
We locally impose a physical upper bound on the hydraulic aperture:
\begin{equation}
    a_h(x_i, y_i)\leq a_m(x_i, y_i), \quad \forall i
    \label{eq:upper_bound}
\end{equation} 
which reflects the tendency of mechanical aperture maps to overestimate fracture conductivity by providing a strict flow barrier and neglecting roughness (Figure~\ref{fig:apertures_models}A). 

    \item \textbf{Physics-based constraint for upscaling}\\
We derive a geometrically corrected permeability field from the hydraulic aperture:
\begin{equation}
     K_{NN}^{\,a_h}(x,y) = \frac{a_h(x,y)^2}{12} \left(1+\alpha \, \frac{|a_h(x,y) - <a_m>|}{\sigma_{a_m}} \right)
\label{eq:geo-corr}
\end{equation}
where $\alpha$ is a learnable correction factor, and $\sigma_{a_m}$ is the standard deviation of the mechanical aperture. The inferred hydraulic aperture field then satisfies the constraint:
\begin{equation}
    K_{NS} = \frac{1}{|\OmegaD{2}|}\int_\OmegaD{2} K_{NN}^{\,a_h}(x,y) \,dx dy + \xi_m, 
\label{eq:phys_const}
\end{equation}
where $ \xi_m\sim \mathcal{N}(0, \sigma_m^2 I)$ and $\sigma_m$ is unknown and accounts for modelling uncertainties. Here, $\xi_m$ capture both model misspecification and numerical errors in the evaluation of $K_{NS}$.     
\end{enumerate}

\vspace{-0.3cm}
The physics-based constraint is designed to correct the local applicability of the Cubic Law, providing a proxy for the hydraulic aperture field that incorporates roughness effects and accurately captures upscaled flow behaviours. For each subvolume, the effective permeability $K_{NS}$ is computed via DNS prior to training and used as an input of the B-PINN correction (see Figure \ref{fig:workflow}). The physical correction leverages the initial distribution of the mechanical aperture field and automatically accounts for the deviation of the inferred hydraulic aperture field from the mean value $<a_m>$ relative to the standard deviation $\sigma_{a_m}$ (see histogram in Figure \ref{fig:workflow}). Equation \eqref{eq:geo-corr} thus guarantees that when the hydraulic aperture is locally close to the mean mechanical aperture, we recover a local permeability that closely aligns with the Cubic Law approximation. Conversely, when the hydraulic aperture falls within the tails of the mechanical aperture distribution, causing significant deviations from the mean value $<a_m>$, the relative roughness term becomes predominant and, therefore, leads to an automatic and local correction of the Cubic Law approximation. We refer to this modelling adjustment as a local relative roughness correction of the Cubic Law. Finally, equation \eqref{eq:phys_const} ensures the arithmetic mean of the local permeability field $K_{NN}^{\,a_h}(x,y)$, derived from the geometrical correction, fits the upscaled Stokes permeability $K_{NS}$. The unknown correction factor $\alpha$ and uncertainty parameters $\sigma_d$, $\sigma_m$ are inferred jointly and adaptively during the sampling. In practice, this integral formulation is approximated by its discrete mean over observation points during inference. A schematic diagram summarising the workflow is presented in Figure \ref{fig:workflow} and further details about the Bayesian formulation are provided in Texts S3 and S4 in SI.

\begin{figure*}[ht!]
\centering
\includegraphics[width=\linewidth]{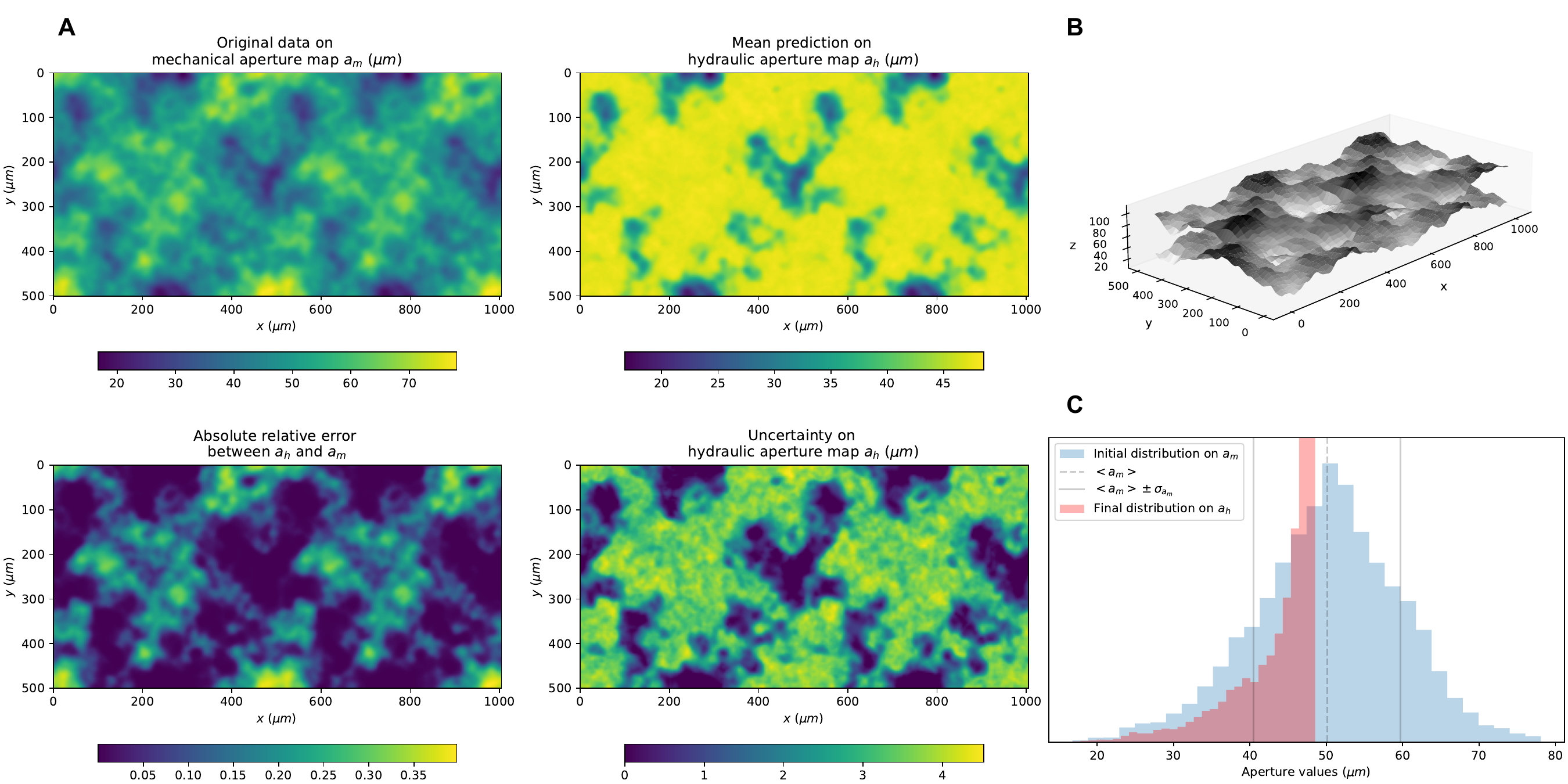}
\caption{Hydraulic aperture inference on rough fracture ($\text{JRC} = 10.31$). (A) 2D maps comparing the original mechanical aperture and mean prediction on the hydraulic aperture (top row), along with the corresponding absolute relative error and local uncertainties on the hydraulic aperture map (bottom row). (B) 3D fracture geometry $\OmegaD{3}$ highlighting the roughness and fracture dimensions in micrometers. (C) Histograms of the aperture values within the fracture domain: initial mechanical aperture distribution (in blue) and final distribution on the mean hydraulic aperture (in red). The dashed line corresponds to the mean mechanical aperture value $<a_m>$, and the two vertical lines delimits the 68\% confidence interval given the standard deviation $\sigma_{a_m}$.} 
\label{fig:hydro_ap}
\end{figure*}

\begin{figure*}[h!]
\centering
\includegraphics[width=\linewidth]{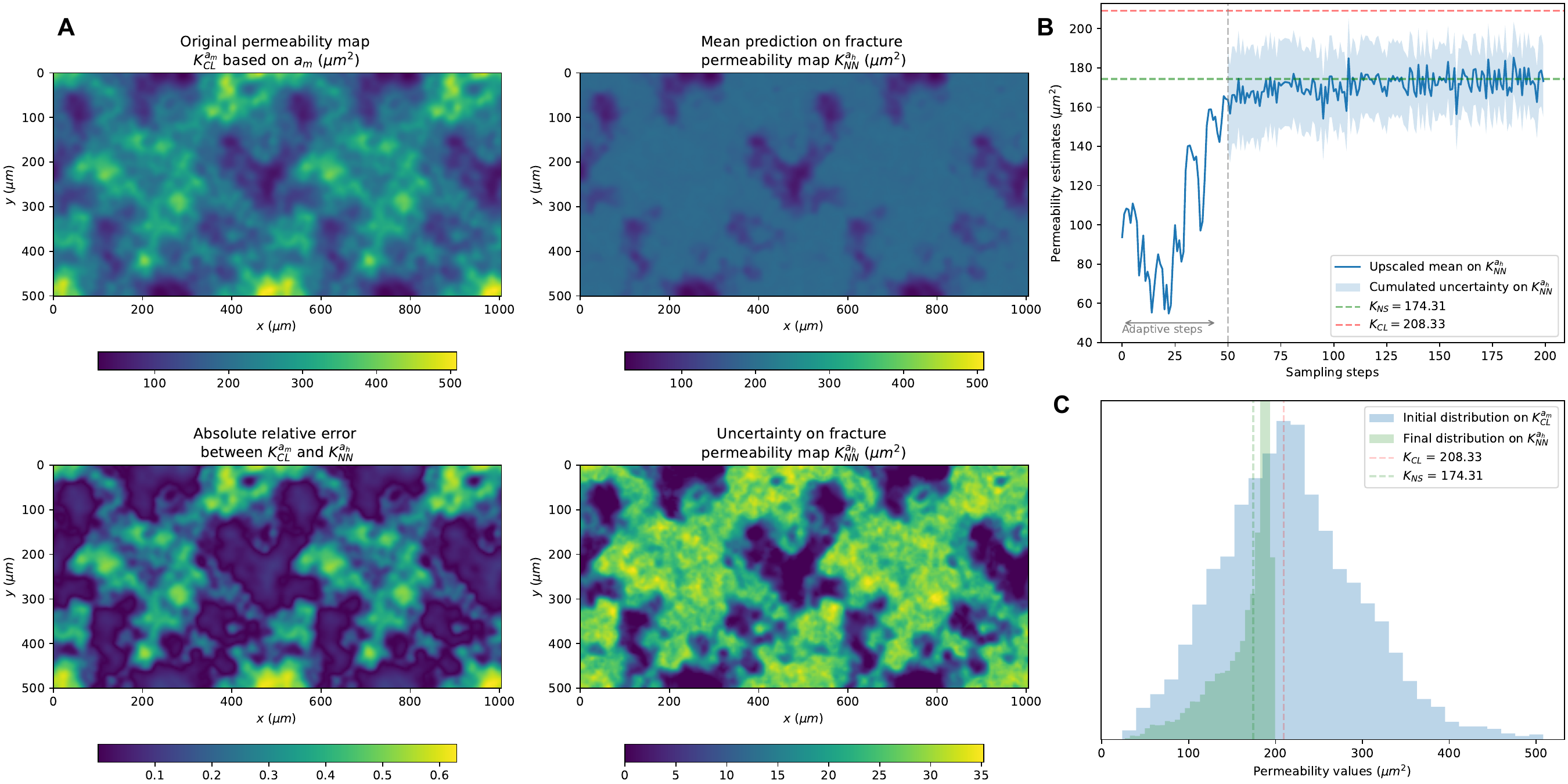}
\caption{Permeability inference on rough fracture ($\text{JRC} = 10.31$). (A) 2D maps comparing the local Cubic Law permeability $K_{CL}^{\,a_m}(x,y)$ and the mean prediction on the local permeability $K_{NN}^{\,a_h}(x,y)$ (top row), along with the corresponding absolute relative error and local uncertainties on the permeability map (bottom row). (B) Convergence of the upscaled permeability based on $K_{NN}^{\,a_h}(x,y)$, together with its uncertainty, toward the unbiased estimate $K_{NS}$ along the inference. The grey vertical dashed line delimits the number of adaptive steps in the AW-HMC sampler. (C) Histograms of the permeability values within the fracture domain: initial distribution on $K_{CL}^{\,a_m}(x,y)$ (in blue) and final distribution on the mean permeability $K_{NN}^{\,a_h}(x,y)$ (in green). The red and green dashed lines correspond, respectively, to biased Cubic Law $K_{CL}$ and effective permeability $K_{NS}$. } 
\label{fig:perm}
\end{figure*}

\section{Results}

We apply our Bayesian correction to the synthetic fracture geometries with increasing wall roughness, defined by their respective JRC values. We also validate our methodology on a smooth, idealised fracture consisting of two parallel planes with a constant mean aperture of $50\,\mu m$. In this theoretical case, all modelling approaches ---  Cubic Law, Darcy flow-based upscaling, Stokes, and B-PINN correction --- provide similar results, as the mechanical and hydraulic apertures closely align. As roughness increases, however, significant discrepancies arise. Figure \ref{fig:hydro_ap} and \ref{fig:perm} present results for the roughest case ($\text{JRC} = 10.31$), whose 3D geometry is illustrated in Figure \ref{fig:hydro_ap}B. In Figure \ref{fig:hydro_ap}A, we compare local maps of the initial mechanical aperture $a_m(x,y)$ with the mean prediction of the hydraulic aperture, obtained by marginalization over the probabilistic posterior distribution. This posterior predictive mean corresponds to an ensemble average across the posterior B-PINN predictions and is defined in Text S3 in SI. Similarly, local uncertainties in the hydraulic aperture are quantified as the standard deviation of the posterior distribution, as shown in Figure \ref{fig:hydro_ap}A. 

These results underscore that regions with the largest mechanical aperture values contribute minimally to effective permeability and upscaled flow behaviour. The inferred hydraulic field exhibits a distinct cut-off in these high-aperture zones, indicating their limited contribution to flow resistance when roughness is substantial. Posterior uncertainties are also concentrated in these regions, which is consistent with the largest deviations observed between $a_h(x,y)$ and $a_m(x,y)$. Although relative errors appear significant in areas of the largest mechanical aperture, this outcome is expected: our approach jointly compensates for data and model errors by combining data-driven and physics-based methods. In particular, we demonstrate that the mechanical aperture field is not a suitable dataset for providing reliable fracture conductivity, as the data-fitting constraint emerges as the most uncertain objective in the Bayesian inference. The AW-HMC sampler automatically detects this initial imbalance, inferring $\sigma_m \leq \sigma_d$. Specifically, the aperture histograms in Figure \ref{fig:hydro_ap}C reveal an automatic distribution shift towards the smallest aperture values, which restrict flow rate and more accurately capture the upscaled hydraulic response.

Beyond reconstructing hydraulic aperture fields, our approach derives permeability maps, based on $a_h(x,y)$, that incorporate roughness effects via local geometric corrections of the Cubic Law. Figure \ref{fig:perm}A presents the posterior predictive mean and associated uncertainties for the corrected permeability field $K_{NN}^{\,a_h}(x,y)$ (see Text S6 in SI for detailed computations). The results are compared against the pointwise permeability map $K_{CL}^{\,a_m}(x,y)$ derived from the local Cubic Law. This shows a significant reduction in permeability in regions of high mechanical aperture, where deviations between $K_{NN}^{\,a_h}(x,y)$ and $K_{CL}^{\,a_m}(x,y)$ are most prominent. Figure \ref{fig:perm}C identifies a distribution shift in the permeability values across the fracture domain: the initial permeability distribution resulting from $K_{CL}^{\,a_m}(x,y)$ is compared with the final distribution on $K_{NN}^{\,a_h}(x,y)$. We observe that the Cubic Law approximation $K_{CL}$ (red dashed line) lies outside the final permeability distribution, while the effective permeability $K_{NS}$ is accurately recovered. These results confirm that our strategy effectively corrects model misspecification and captures small-scale uncertainties through a well-balanced combination of data-driven and physics-based constraints. The latter is reinforced by the convergence of the sampler towards the Pareto-optimal front, as well as the convergence of the upscaled permeability values derived from $K_{NN}^{\,a_h}(x,y)$. Indeed, we produce a local permeability field that satisfies the physics-based constraint \eqref{eq:phys_const} in the sense that its arithmetic mean converges towards the Stokes estimate $K_{NS}$ rather than the Cubic Law permeability $K_{CL}$, as shown in Figure \ref{fig:perm}B. Notably, the methodology is independent of the choice of the arithmetic mean in equation \eqref{eq:phys_const} and yields comparable results with the geometric or harmonic means.

Figure \ref{fig:apertures_models}B compares the upscaled permeability values from the posterior predictive mean of $K_{NN}^{\,a_h}(x,y)$ --- as defined in Equation S13 in SI --- with those from other modelling approaches across the four fracture geometries. We also report corresponding uncertainty ranges for the upscaled permeability values, which are derived from the local uncertainties in the permeability field predictions (see Equation S15 in SI). These results demonstrate that our AI-driven correction framework reliably recovers mean upscaled permeability and uncertainty bounds, capturing roughness-induced effects in fracture conductivity estimation. We successfully generate probabilistic permeability maps that are compatible (by construction) with Stokes upscaling for each fracture geometry. Notably, these corrected maps are now suitable and relevant for 2D Darcy flow-based upscaling, which, in turn, yields results compatible with Stokes upscaling (see Text S6 and Table S2 in SI). In other words, our approach recovers a local permeability field that ensures robust estimation of fracture conductivity through Darcy upscaling, by leveraging the latent hydraulic aperture rather than relying on error-prone measurements of the mechanical aperture. This establishes a robust framework for estimating uncertainty ranges on conductivity at the larger scale, supporting the extension of our methodology to complex fracture networks.

\section{Discussion and Conclusion}

We introduce an AI-driven uncertainty quantification framework that mitigates bias due to model misspecification in fracture conductivity and provides transparent estimates of the resulting uncertainties. This probabilistic approach represents a crucial step toward understanding how small-scale uncertainties propagate and influence fracture network conductivity at larger scales. By delivering more realistic ranges of network hydraulic conductivity, we support robust calibration of fault zone properties for sensitivity analyses at the reservoir-caprock scale, which are critical for assessing fault-related leakage and de-risking CCS facilities. Although developed for \ce{CO2} storage, our approach is broadly relevant and applicable to other subsurface systems where fractures govern flow, including geothermal energy production, and groundwater management in fractured aquifers. 

We demonstrate that common approximations relying on measurements of the mechanical aperture, such as the empirical Cubic Law and Darcy upscaling, fail to capture the roughness effects and smaller-scale uncertainties. These modelling misspecifications lead to systematic overestimation of fracture conductivity, significantly impacting the upscaled flow behaviour of realistic fracture networks.

Therefore, we propose a deep learning alternative that learns the mapping between the unreliable mechanical and latent hydraulic aperture maps, which captures the small-scale flow behaviours and uncertainties. This is achieved through a multi-objective Bayesian inference that combines data-driven constraints with a physics-based regularisation, correcting the local Cubic Law limitations by accounting for the roughness. This B-PINN framework, powered by the AW-HMC sampler, adaptively characterises modelling and data uncertainties, enabling robust inference and efficient exploration of the optimal Pareto front. The inferred hydraulic aperture distributions reveal a reduced influence of the largest mechanical aperture regions on fluid flow in rough fractures. We further generate corrected permeability maps based on the hydraulic apertures, ensuring their upscaling effectively captures the unbiased Stokes estimate of permeability. Notably, these probabilistic permeability maps are now Darcy-consistent and can serve as inputs for flow-based upscaling, enabling accurate and uncertainty-aware modelling of conductivity across various roughness levels and scales.

This framework paves the way for a multi-scale, AI-driven uncertainty propagation workflow in complex fracture networks, based on pattern recognition of mechanical aperture maps. The objective is to conduct Bayesian inference, as introduced in this work, on selected extracted subvolumes of the fracture network. These subvolumes should be representative of different roughness patterns and aperture distributions; therefore, their appropriate resolutions can be determined through prior statistical analysis of mechanical aperture histograms or correlation lengths. Such a dataset consisting of paired images of mechanical apertures $a_m(x,y)$ and probabilistic functions of the permeability field $K_{NN}^{\,a_h}(x,y)$, evaluated on the extracted subvolumes, can subsequently be input into purely data-driven machine learning models. This will lead to improved generalisation to previously unseen fracture network geometries, while simultaneously enhancing the propagation of uncertainties across different scales. This approach will then yield a complete corrected permeability map for the entire 3D fracture network that locally accounts for the geometric characteristics of individual fractures at the small scale, and can finally be coupled with Darcy flow-based upscaling at the larger scale. Similar techniques, relying on Convolutional Neural Networks, have successfully been employed to estimate the upscaled permeability of rock cores from $\mu$CT images \cite{Jiang_2023}, as well as other properties of porous media such as porosity or specific surface area \cite{ALQAHTANI2020}. The primary advantage is the low computational cost associated with Bayesian inference, which can be executed in parallel across various subvolumes. Furthermore, this will enhance the generalisation of the AI workflow to different fracture network geometries, rendering it an attractive alternative to DNS methods. Upscaling the uncertainties in fracture network conductivity is also of crucial importance when it comes to ensuring reliable estimates of fault-leakage rates. Overall, such a deep learning based upscaling will allow estimating the conductivity of a complex fracture network, with large field of view, while preserving the effects of small-scales uncertainties in terms of the local heterogeneities, geometric variations and roughness.

Finally, our method will support future investigation into uncertainty propagation in multi-phase flow processes. Roughness and aperture variability significantly influence capillary trapping and fluid phase distributions in geological fractures \cite{YANG2016, Guiltinan_2021, He_2022, Ma2024, PNAS_2023_Two_phase_rough}. Similarly, leveraging a reliable mapping between mechanical aperture and locally corrected permeability fields is essential for understanding how small-scale uncertainties influence hydro-mechanical coupled processes \cite{CUNHA2024111094}. Investigating the uncertainties associated with these processes remains critical in the context of fault-related leakage for \ce{CO2} storage in the subsurface.

\section*{Supporting Information}

\setcounter{figure}{0}
\renewcommand{\figurename}{Fig.}
\renewcommand{\thefigure}{S\arabic{figure}}
\setcounter{table}{0}
\renewcommand{\thetable}{S\arabic{table}}

\noindent\textbf{Text S1. Stokes Permeability} \\

The effective absolute permeability $K_{NS}$, used throughout this study (see Fig.~2 in the main text), is computed via Direct Numerical Simulation of incompressible Stokes flow within the realistic 3D fracture geometry $\Omega^{3D}$. The domain is binary decomposed into solid and fluid subdomains, $\Omega^{3D}_s$ and $\Omega^{3D}_f$, separated by the fracture interface $\Sigma$. We assume impermeable fracture walls and neglect contribution of the surrounding porous matrix, which is consistent with fault-related leakage models where fracture flow dominates due to significantly lower matrix permeability.

Simulations are performed along the primary flow direction (x-axis), assuming low Reynolds regime and thus incompressible Stokes equation with adherent boundary conditions on the fracture walls:
\begin{equation}
\left\{
\begin{array}{ll}
    -\mu \Delta u +\nabla p = f & \quad \text{in } \Omega^{3D}_f \\[1mm]
    \nabla \cdot u = 0 & \quad \text{in } \Omega^{3D}_f \\[1mm]
    u = 0 & \quad \text{on } \Sigma \\[1mm]
\end{array}
\right.
\label{Stokes}
\end{equation}
where $u$ is the pore-scale velocity, $p$ the pressure, $\mu$ the dynamic viscosity set to $\mu = 1cP$, and $f$ the body force. Periodicity is imposed on the inlet and outlet boundaries via thin fluid layers. In practice, the system is solved in a dimensionless form to improve numerical stability, using an equivalent driving force of $f = 10\,Pa.m^{-1}$.

The effective permeability is computed by upscaling the simulated flow response to the Darcy scale:
\begin{equation}
K_{NS} = \frac{\phi \mu \langle u_x \rangle_{\Omega^{3D}_f}}{\langle f_x \rangle_{\Omega^{3D}_f}},
\label{Kdef}
\end{equation}
where $u_x$ and $f_x$ are the x-components of the velocity and driving force, $<.>_{\Omega^{3D}_f}$ represents the average in the fluid domain, $\phi$ the fracture macro-porosity and $v = \phi<u>_{\Omega^{3D}_f}$ is the so-called superficial or Darcy velocity \cite{QW88}. The porosity is set to $\phi = 1$ to isolate the fracture contribution, assuming negligible matrix flow.

\begin{figure}
\centering
\includegraphics[width=0.5\textwidth]{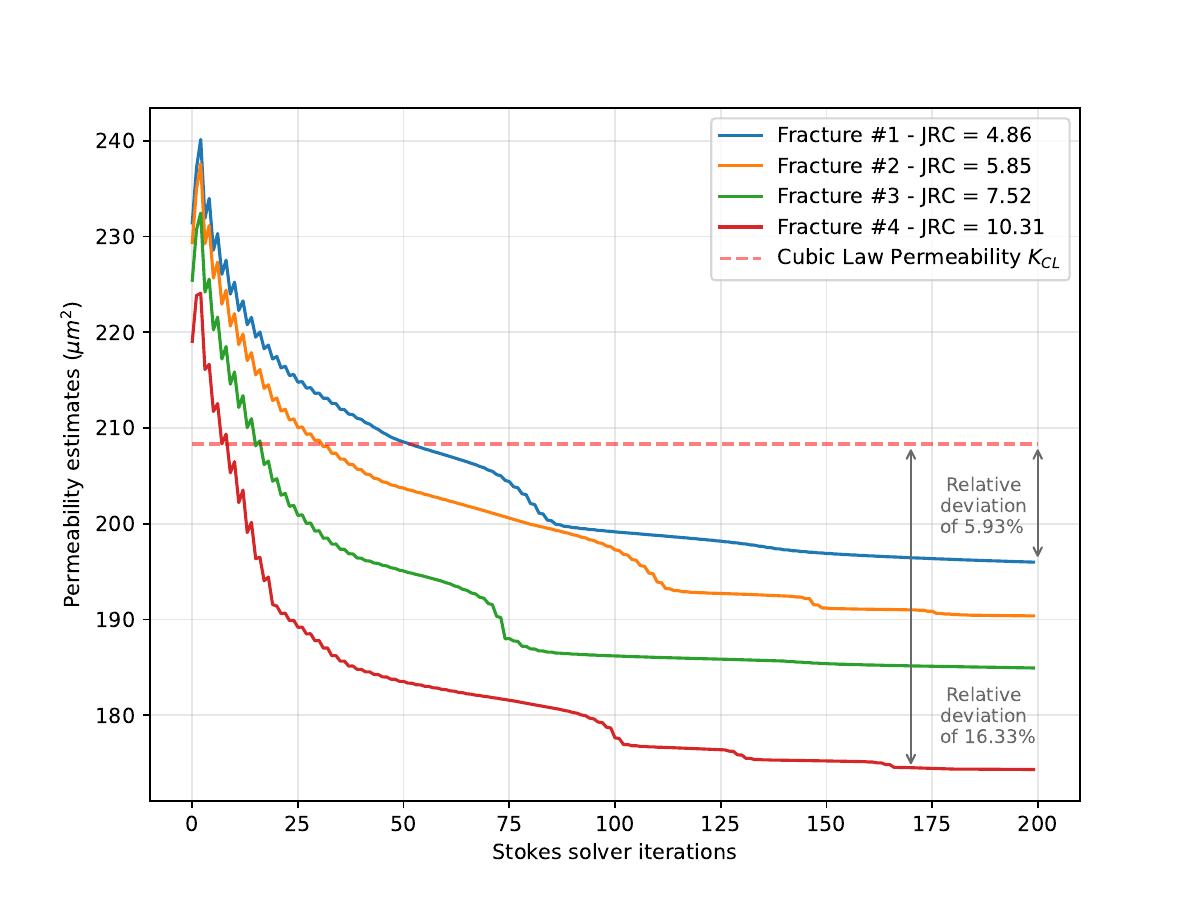}
\caption{Evolution of permeability estimates over the iterations of the Stokes solver for the different fracture geometries, with roughness characterised by the JRCs. The dashed horizontal line represents the overestimated Cubic Law approximation. We observe deviations in fracture permeability due to roughness, ranging from approximately $6\%$ to $16\%$. The estimations of Stokes permeabilities for the different JRCs are performed prior to the Bayesian inference.}
\label{fig:perm_conv}
\end{figure}

Flow simulations are performed using the DNS Stokes solver developed by Perez et al. \cite{Perez2022} which discretizes the Stokes equations into a linear system and iteratively solves for the velocity and pressure fields using a Krylov subspace method (GMRES), until a residual-based convergence criterion is met. The 3D computational domain closely aligns with the fracture geometries and consists of a grid with a resolution of $256\times128\times30$, scaled such that $\delta x = 3.94\,\mu m$. Figure~\ref{fig:perm_conv} shows permeability convergence and relative deviation from the Cubic Law approximation $K_{CL}$, computed as $(K_{CL} - K_{NS}) / K_{CL}$, across the four synthetic fractures with varying roughness. Deviations range from $6\%$ to $16\%$ for the different JRC coefficients and the final effective permeability values are summarised in Table S1. A sub-1\% deviation is observed in the smooth case ($\text{JRC} = 0$), likely due to minor meshing inaccuracy ($\langle a_m \rangle \simeq 50.19\,\mu m$).

\begin{figure*}
\centering
\includegraphics[width=\textwidth, trim={1.3cm 0cm 2.5cm 1cm},clip]{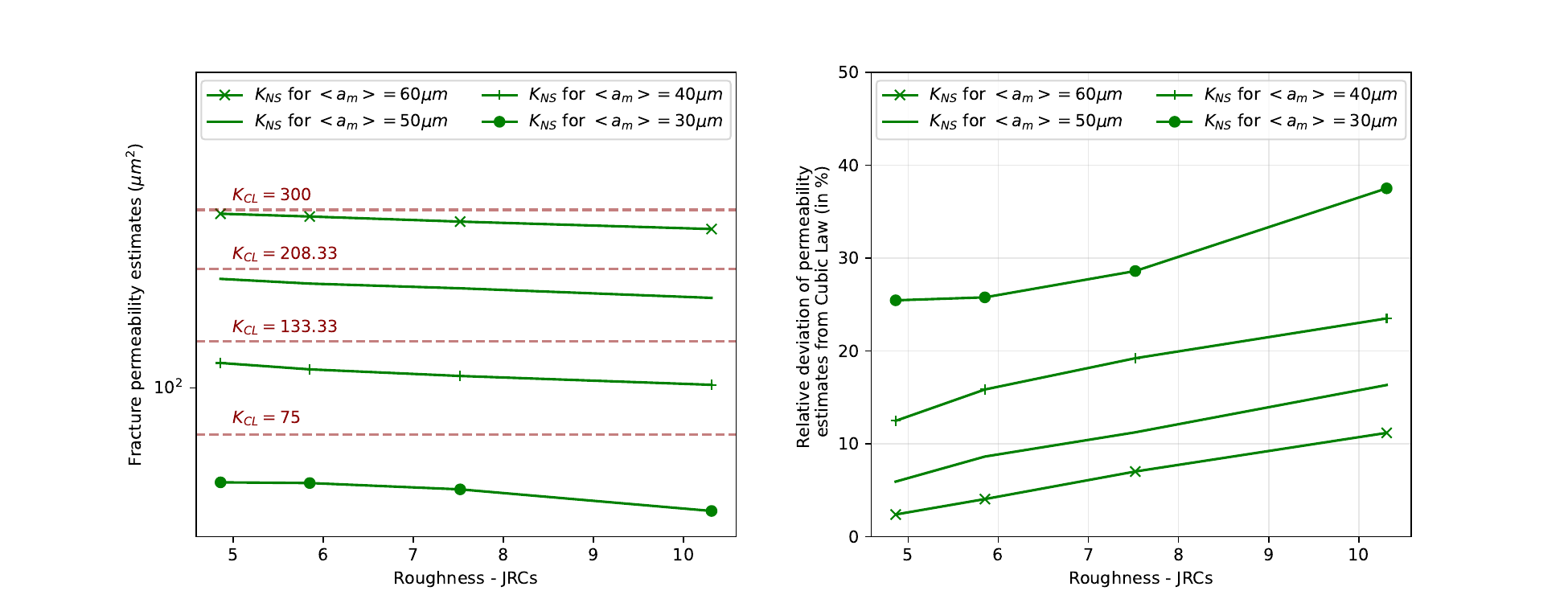}
\caption{Comparison of Stokes permeability estimates for various fracture geometries, with roughness characterized by JRC values and mean apertures ranging from $30\mu m $ to $60 \mu m$ (on the left). For each JRC, the fracture geometries remain identical, while the mean aperture varies relatively to the reference case ($<a_m> = 50\mu m$). Dashed horizontal lines represent the overestimated fracture permeability obtained from direct Cubic Law approximations for each mean mechanical aperture value $<a_m>$. Relative deviations in permeability with respect to the Cubic Law, computed as $(K_{CL}-K_{NS})/K_{CL}$, for different JRCs and varying mean apertures (on the right). This highlights greater deviations for smaller aperture values, reaching nearly 40\%, and emphasizes the importance of considering corrections for such model misspecifications.}
\label{fig:mean_ap_effect}
\end{figure*}

We also investigate the effect of varying the mean mechanical aperture (from $30$ to $60\,\mu m$) while preserving the fracture geometry for each JRCs. Results (see Fig.~\ref{fig:mean_ap_effect}) reveal that deviations from the Cubic Law increase for smaller apertures: up to 25\% for $\text{JRC} = 4.86$ and nearly 40\% for $\text{JRC} = 10.31$. These results confirm the non-negligible impact of fracture roughness and mean mechanical aperture on the estimation of fracture conductivity. \\

\noindent\textbf{Text S2. 2D Darcy Flow-Based Upscaling} \\

Darcy flow-based upscaling of fracture permeability is performed using the MATLAB Reservoir Simulation Toolbox (MRST) for single-phase flow \cite{Lie_2019}. The initial permeability distribution is computed by applying the local Cubic Law to the mechanical aperture field $a_m(x,y)$, yielding the pointwise permeability map $K_{CL}^{\,a_m}(x,y)$ (see Eq.~3 and Fig.~4A in the main text). The grid resolution of $256 \times 128$ is strictly identical to the one used for extracting the training dataset $\mathcal{D}$ in the Bayesian inference.

The Darcy flow solver approximates the solution of the single-phase pressure equation:
\begin{equation}
   - \nabla \cdot \left( \frac{K_{CL}^{\,a_m}(x,y)}{\mu} \nabla p \right) = 0,
   \label{eq:Poisson}
\end{equation}
where $v = - K_{CL}^{\,a_m}(x,y) \mu^{-1} \nabla p$ is the Darcy velocity, $\mu$ is the dynamic viscosity, and $p$ is pressure. Fixed pressure boundary conditions are applied at the inlet and outlet, replicating the driving force used in Stokes simulations, while no-flow conditions are applied on the remaining boundaries.

The upscaled permeability is obtained from the average outflow velocity $q$, computed as the ratio of the flow rate $Q$ (in $m^3.s^{-1}$) by the cross sectional area $A$ (in $m^2$), and the pressure drop $\Delta p$ over the domain length $L_x$:
\begin{equation}
   K_D^{\,a_m} = \frac{\mu q L_x}{\Delta p}.
\end{equation}

These estimates are compared to the global Cubic Law $K_{CL}$ and the effective Stokes permeability $K_{NS}$ (Table S1) and underscore that Darcy upscaling based on the mechanical aperture measurements is not sufficient to capture intrinsic roughness effects. \\

\noindent\textbf{Text S3. Posterior predictive mean and uncertainty} \\

The B-PINN formulation seeks to infer the probabilistic posterior distribution of a set of unknown parameters, $\Theta = \{\theta, \mathcal{P}_{inv}\}$ which includes the Bayesian neural network parameters $\theta$ and potential inverse parameters $\mathcal{P}_{inv}$ associated with the physical model. The posterior distribution combines a data likelihood term $P(\mathcal{D}|\Theta)$, assessing the fit to the observations $\mathcal{D}$, a physics-based likelihood term $P(\mathcal{M}|\Theta)$, accounting for model $\mathcal{M}$ discrepancies and misspecifications, and a joint prior distribution $P(\Theta)$. This results in the sampling of a high-dimensional and multi-task posterior distribution over the parameters $\Theta$, expressed as: 
\begin{equation}    
    P(\Theta|\mathcal{D}, \mathcal{M}) \propto P(\mathcal{D}|\Theta)P(\mathcal{M}|\Theta)P(\Theta).
\label{eq:post_dist_theta}
\end{equation}

The distribution of solutions, which are close to the optimum and characterise the uncertainties of the problem, is obtained through a marginalisation over the posterior distribution of B-PINN predictions \cite{wilson_bayesian_2020}. The latter translates the posterior distribution over the parameters $\Theta$ into a probability function over the fields of interest. Specifically, we are interested in quantifying
\begin{equation}
    P(a_h|X, \mathcal{D}, \mathcal{M}) = \int P(a_h|X,\Theta) P(\Theta|\mathcal{D}, \mathcal{M}) d\Theta
\label{eq:BMA}
\end{equation}
where $X$ are the neural network inputs (\emph{e.g.} spatial coordinates in $\OmegaD{2}$) and $a_h$ is the predictive output of the B-PINNS representing the hydraulic aperture field. Equation \eqref{eq:BMA} defines the posterior predictive distribution, obtained by averaging predictions over the parameter posterior \eqref{eq:post_dist_theta}, and ultimately provides a mean prediction (the posterior predictive mean) along with uncertainties on the hydraulic aperture field.

In practice, the posterior predictive mean is computed as the expectation of \eqref{eq:BMA}, which we approximate by Monte Carlo averaging over posterior samples:
\begin{equation}
    \mathbb{E}[a_h \,|\, X, \mathcal{D}, \mathcal{M}]
    \simeq \frac{1}{N_s - N} \sum_{i=N}^{N_s} a_h(X;\Theta^{t_i})
    \label{eq:BMA_approx}
\end{equation}
where $a_h(X;\Theta^{t_i})$ represents the surrogate model prediction of the hydraulic aperture, evaluated at the spatial coordinates $X = \{(x,y) \in \Omega^2\}$, and resulting from the sampling iteration $i$ of the set of parameters $\Theta$. In equation \eqref{eq:BMA_approx}, $N_s$ refers to the total number of sampling steps and $N$ is the number of adaptive steps in the AW-HMC sampler, such that the samples $\left\{\Theta^{t_i}\right\}_{i=N}^{N_s}$ are theoretically drawn from the target posterior distribution $\Theta^{t_i} \sim P(\Theta| \mathcal{D}, \mathcal{M})$. The reader is referred to \cite{Perez2023} for theoretical background, methodological and algorithmic details on the AW-HMC sampler.\\

\begin{table*}[ht!]
\centering
\begin{tabular}{cccccc}
\hline 
Roughness & Cubic Law & Darcy & Stokes & \multicolumn{2}{c}{Bayesian PINN} \\
$\text{JRC}$ & $K_{CL}$ & $K_{D}^{\,a_m}$ & $K_{NS}$ & Mean on $K_{NN}^{\,a_h}$ &  UQ on $K_{NN}^{\,a_h}$\\
\hline 
/ & $\mu m^2$ & $\mu m^2$ & $\mu m^2$ &$\mu m^2$ &$\mu m^2$ \\
\hline
    $0$ & \multirow{5}{*}{\textcolor{darkred}{$208.33$}} & \textcolor{seagreen}{$208.33$} & \textcolor{seagreen}{$209.94$} & \textcolor{seagreen}{$208.33$} & \textcolor{seagreen}{/} \\
$4.86$ & & \textcolor{darkred}{$201.98$} & \textcolor{seagreen}{$195.98$} & \textcolor{seagreen}{$189$} & \textcolor{seagreen}{$[178;199]$} \\
$5.85$ & & \textcolor{darkred}{$201.75$} & \textcolor{seagreen}{$190.36$} & \textcolor{seagreen}{$191$} & \textcolor{seagreen}{$[184;199]$} \\
$7.52$ & & \textcolor{darkred}{$201.61$} & \textcolor{seagreen}{$184.92$} & \textcolor{seagreen}{$187$} & \textcolor{seagreen}{$[177;196]$} \\
$10.31$ & & \textcolor{darkred}{$201.55$} & \textcolor{seagreen}{$174.31$} & \textcolor{seagreen}{$171$} & \textcolor{seagreen}{$[153;189]$} \\
\hline
\end{tabular}
\vspace{0.5cm}
\caption{Comparison of permeability estimates from different modelling assumptions across various roughness patterns, defined by their JRC values, for the four fracture geometries of the dataset. The case with no roughness ($\text{JRC} = 0$) is also considered.}
\label{tableS1}
\end{table*}

\noindent\textbf{Test S4. A Multi-Objective Bayesian Inference Problem}\\

Exploring the uncertainties related to inferring the latent hydraulic aperture field $a_h(x,y)$ entails solving a multi-objective problem that integrates data with modelling constraints in a probabilistic framework. The data constraint first leverages measurements of the mechanical aperture such that $a_m(x_i,y_i) = a_h(x_i, y_i) + \xi_d(x_i, y_i)$, with $\xi_d\sim \mathcal{N}(0, \sigma_d^2I)$ and the standard deviation $\sigma_d^2$ unknown. Given the AW-HMC sampler, this leads to the following expression of the data-fitting likelihood term in equation \eqref{eq:post_dist_theta}:
\begin{equation}
    P(\mathcal{D}|\Theta) \propto \mathrm{exp}\left(-\frac{1}{2\sigma_d^2} \|a_h - a_m\|_\mathcal{D}^2\right)
\label{eq:data_likelihood}
\end{equation}
which quantifies the discrepancies between the mechanical aperture measurements and the inferred hydraulic apertures. The inferred hydraulic aperture field should also satisfy the constraint 
\begin{equation}
    K_{NS} = \frac{1}{|\OmegaD{2}|}\int_\OmegaD{2} K_{NN}^{\,a_h}(x,y) \,dx dy + \xi_m
\end{equation}
with $\xi_m\sim \mathcal{N}(0, \sigma_m^2 I)$ and $\sigma_m$ unknown. The latter translates into the physics-based likelihood in equation \eqref{eq:post_dist_theta} given by 
\begin{equation}
   \scriptstyle P(\mathcal{M}|\Theta) \propto \mathrm{exp}\left(-\frac{1}{2\sigma_m^2} \left\|K_{NS} - \frac{1}{|\OmegaD{2}|}\int_\OmegaD{2} K_{NN}^{\,a_h}(x,y) \,dx dy \right\|_\mathcal{D}^2\right)
\label{eq:phys_likelihood}
\end{equation}
with the correction factor $\alpha$ adaptively and automatically inferred during the sampling. 

The values $K_{NS}$ are computed prior to the Bayesian inference for each synthetic fracture or subvolume in the context of large fracture networks. Notably, local computations of $K_{NS}$ on multiple extracted subvolumes are computationally more efficient compared to solving Stokes across the entire complex 3D fracture network. Therefore, this approach will allow for fast upscaling to realistic fracture network geometries by leveraging several geometrical and probabilistic mappings between $a_m$, $a_h$ and $K_{NN}^{\,a_h}$. It can subsequently provide reliable estimates of the conductivity of fracture networks while accounting for smaller-scale uncertainties due to roughness. 

In practice, the integral formulation used in the physics-based likelihood constraint \eqref{eq:phys_likelihood} is approximated by its discrete arithmetic mean over the $N_{obs}$ observation points during the sampling phase of the B-PINNs. In contrast, during the prediction phase, the validity of the arithmetic upscaling is evaluated over the entire domain $\OmegaD{2}$, leading to reliable predictions of the hydraulic aperture and local corrected permeability fields on $\OmegaD{2}$.\\

\noindent\textbf{Text S5. AW-HMC for Multi-Objective B-PINNs} \\

In the B-PINN framework, the multi-objective structure of the posterior distribution \eqref{eq:post_dist_theta} is naturally expressed as a weighted sum of potential energy components $U_k(\Theta)$, forming the total potential energy $U(\Theta)$ of a conservative Hamiltonian system:
\begin{equation}
    U(\Theta) = \sum_{k=1}^{K+1} \lambda_k U_k(\Theta).
\end{equation}
where each $U_k(\Theta)$ corresponds to a distinct objective.
For instance, the data-fitting likelihood follows the distribution $ P(\mathcal{D}|\Theta) \propto e^{-\lambda_1U_1(\Theta)}$ and each component of the multi-objective posterior distribution can be expressed in relation to the components of the weighted multi-potential energy $U(\Theta)$. The weighting parameters $\lambda_k$ represent the relative importance and uncertainty associated with each objective \cite{Perez2023}. This formulation allows the posterior to explicitly account for multiple sources of uncertainty, including data noise, model discrepancy, and prior regularisation.

\begin{figure*}[ht!]
\centering
\includegraphics[width=0.8\textwidth,trim={1.3cm 1.2cm 2.2cm 1.5cm},clip]{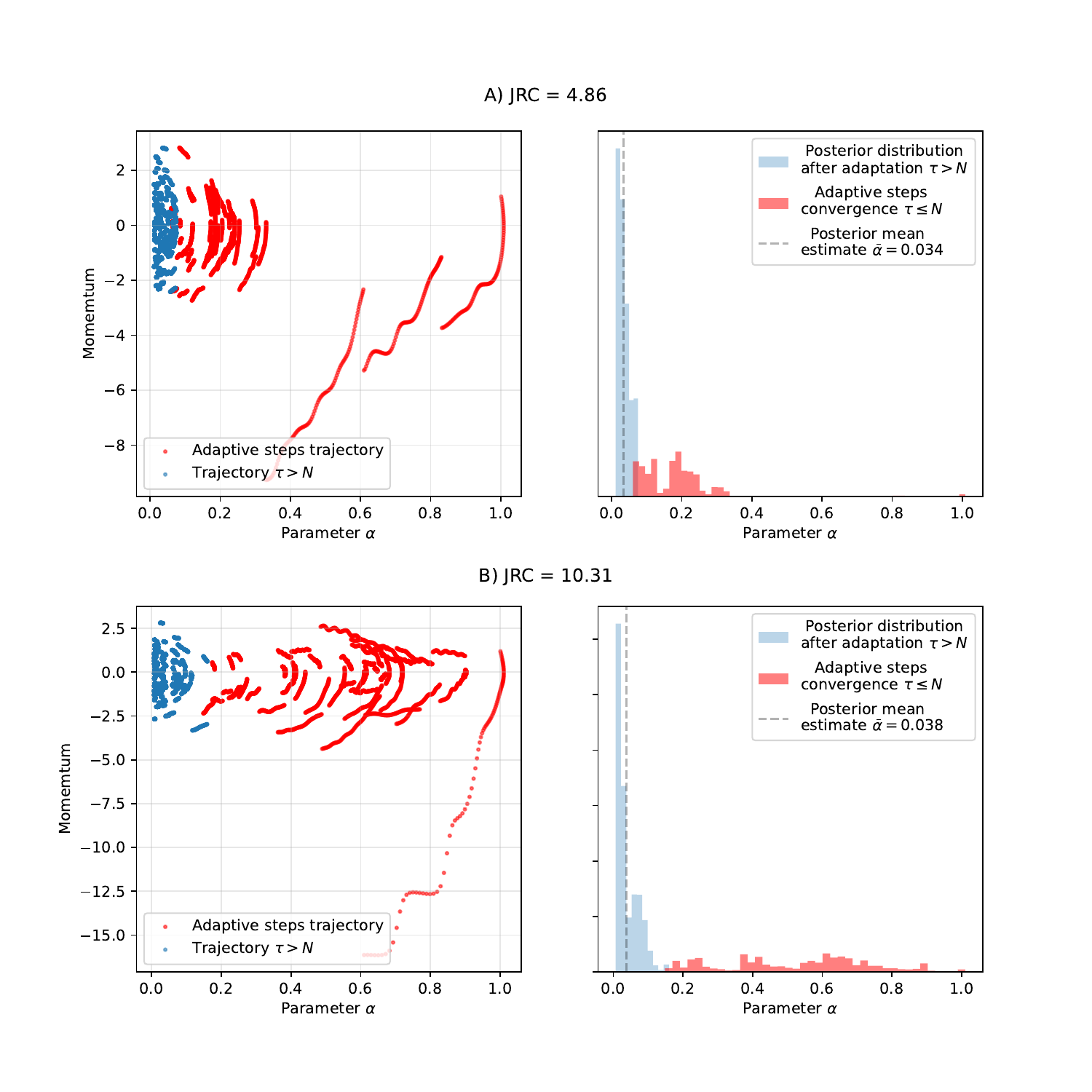}
\caption{Inference of the inverse parameter $\alpha$, which serves as the correction factor in the physics-based constraint for the rough fractures \#1 and \#4 ($\text{JRC} = 4.86$ and $10.31$). Phase diagram of the inverse parameter trajectory during sampling, with the adaptive steps trajectory (in red) and the effective sampling (in blue), on the left. Histogram of the marginal posterior distributions of the parameters $\alpha$, illustrating the distribution tail (in red) due to convergence during the adaptation and the final posterior distribution (in blue), on the right. The posterior mean estimates $\bar\alpha$ are derived from the samples collected after the adaptation process. We obtain a log-normal posterior distribution, ensuring the positivity of the inverse parameters $\alpha$.}
\label{fig:alpha_inf}
\end{figure*}

The AW-HMC sampler aims to automatically manage the different task uncertainties through an adaptive weighting of the target posterior distribution based on a control of the weights $\lambda_k$. The gradient variances of the different tasks are leveraged for a number of adaptive steps $\tau\leq N$ to ensure the gradient distributions of each potential energy term $U_k(\Theta)$ have balanced contributions: 
\begin{equation}
\label{Lambda_weights}
    \lambda_k =\left( \frac{\gamma^2}{\mathrm{Var}\{\nabla_\Theta U_k \} }\right)^{1/2}, 
\end{equation}
with \begin{equation*}
    \gamma^2 := \min_{t=1..K} (\mathrm{Var}\{\nabla_\Theta U_t \}),\quad \forall k = 1,...,K.
\end{equation*}
This approach avoids the vanishing of task-specific gradients \cite{Sener_2018, Rahaman_spectral_2019, Maddu_2022} and allows the sampling to focus on the neighbourhood of the Pareto front after the adaptive steps. This ensures unbiased predictions and improves the convergence and stability of the sampler compared to commonly used alternatives, including traditional HMC and No-U-Turn (NUTS) samplers (see \cite{Perez2023} for a full comparison). The last weight $\lambda_{K+1}$, associated with the prior term $P(\Theta)$, is not adjusted in the same manner as outlined in \eqref{Lambda_weights}, because it serves as a regularisation term similar to that used in multi-objective optimisation. In particular, the joint prior distribution on the set of parameters $\Theta$ follows a multivariate Gaussian distribution, assuming independent variables, such that $P(\Theta)\sim \mathcal{N}(0, \sigma^2 I_p)$. However, to ensure the positivity of the inverse parameter, specifically the correction factor $\alpha$, we typically assume a log-normal prior distribution for $\alpha$ and apply the following change of variable: $\alpha = e^{\Tilde{\alpha}}$. This appropriate change of variable allows us to consider Gaussian prior distribution on the newly defined set of parameters $\Theta = \{\theta, \Tilde{\alpha}\}$, with $\theta$ the Bayesian neural network parameters.

Finally, we can obtain a posterior distribution for the unknown correction factor $\alpha$. The phase space trajectories of the parameter $\alpha$ and the histogram of its marginal posterior distribution are presented in Fig. \ref{fig:alpha_inf} for the rough fractures \#1 and \#4, with respectively $\text{JRC} = 4.86$ and $10.31$. This figure shows in both cases the convergence of the parameter towards its mode during the adaptive steps $\tau\leq N$ of the AW-HMC sampler, as indicated in red. Once the adaptive process ends, we begin sampling the posterior distribution of $\alpha$, as shown in blue. The posterior means of this inverse parameter can also be estimated as $\bar\alpha = 3.4\times 10^{-2}$ and $3.8\times 10^{-2}$ for their respective JRCs.\\

\begin{table*}[h!]
\centering
\begin{tabular}{ccccccc}
\hline
Roughness & Original Darcy & \multicolumn{2}{c}{Bayesian PINN}& \multicolumn{2}{c}{Corrected Darcy} \\
$\text{JRC}$ & $K_{D}^{\,a_m}$ & Mean on $K_{NN}^{\,a_h}$ &  UQ on $K_{NN}^{\,a_h}$ & Mean on $K_{D}^{\,a_h}$ &  UQ on $K_{D}^{\,a_h}$\\
\hline
/ & $\mu m^2$ & $\mu m^2$ &$\mu m^2$ &$\mu m^2$&$\mu m^2$ \\
\hline
$4.86$ & \textcolor{darkred}{$201.98$} & \textcolor{seagreen}{$189$} & \textcolor{seagreen}{$[178;199]$}& \textcolor{seagreen}{$184$} & \textcolor{seagreen}{$[174;193]$} \\
$5.85$ & \textcolor{darkred}{$201.75$} & \textcolor{seagreen}{$191$} & \textcolor{seagreen}{$[184;199]$}&\textcolor{seagreen}{$186$} & \textcolor{seagreen}{$[179;192]$} \\
$7.52$ & \textcolor{darkred}{$201.61$} &  \textcolor{seagreen}{$187$} & \textcolor{seagreen}{$[177;196]$} & \textcolor{seagreen}{$182$} & \textcolor{seagreen}{$[174;190]$}\\
$10.31$ & \textcolor{darkred}{$201.55$} & \textcolor{seagreen}{$171$} & \textcolor{seagreen}{$[153;189]$} & \textcolor{seagreen}{$168$} & \textcolor{seagreen}{$[151;184]$}\\
\hline
\end{tabular} 
\vspace{0.5cm}
\caption{Comparison of fracture permeability estimates from different modelling assumptions across various roughness patterns (Continuation). This table complements Table S1 and confirms that, when applied to the corrected permeability maps derived from local hydraulic apertures, 2D Darcy flow-based upscaling provides permeability estimates which are consistent with both Stokes and B-PINN inference.}
\label{tableS2}
\end{table*}

\noindent\textbf{Text S6. Corrected Darcy Upscaling Based on Hydraulic Aperture Inference}\\

Once the Bayesian inference correcting the model misspecifications is performed, we recover the hydraulic aperture $a_h(x,y)$ and corrected permeability $K_{NN}^{\,a_h}(x,y)$ fields for each posterior sample $\Theta^{t_i}$, where $i = N, ..., N_s$. The posterior predictive mean of the permeability $K_{NN}^{\,a_h}(x,y)$ is computed as follows:
\begin{equation}
    \mathbb{E}[K_{NN}^{\,a_h} \,|\, X, \mathcal{D}, \mathcal{M}]
    \simeq \frac{1}{N_s - N} \sum_{i=N}^{N_s} K_{NN}^{\,a_h}(X;\Theta^{t_i})
    \label{eq:Mean_K}
\end{equation}
with
\begin{equation}
    \scriptstyle K_{NN}^{\,a_h}(X;\Theta^{t_i}) = \frac{a_h(X;\Theta^{t_i})^2}{12} \left(1+\alpha_{\Theta^{t_i}} \, \frac{|a_h(X;\Theta^{t_i}) - \langle a_m \rangle|}{\sigma_{a_m}} \right)
    \label{eq:K_def}
\end{equation}
where $a_h(X;\Theta^{t_i})$ and $\alpha_{\Theta^{t_i}}$ represent the surrogate model prediction of the hydraulic aperture and the inferred correction factor at each sampling step (see Fig. \ref{fig:alpha_inf} for its evolution).

The mean permeability field \eqref{eq:Mean_K}, denoted more compactly as $\overline{K_{NN}^{\,a_h}(x,y)}$, is used as input for Darcy upscaling, solving Eq.~\eqref{eq:Poisson} with the permeability tensor $\mathbf{K} = \overline{K_{NN}^{\,a_h}(x,y)}$ and identical boundary conditions. The resulting upscaled and corrected permeability, denoted $K_D^{\,a_h}$, closely matches both the Stokes estimate $K_{NS}$ and the upscaled value obtained from the posterior predictive permeability field, in contrast to the original Darcy upscaling based on the mechanical aperture, $K_D^{\,a_m}$ (see Table \ref{tableS2}). The hierarchical relationships in the modelling of fracture permeability can therefore be expressed as $K_{CL} > K_D^{\,a_m} > K_{NS}\simeq K_{NN}^{\,a_h} \simeq K_{D}^{\,a_h}$.

Uncertainty ranges for $K_{D}^{\,a_h}$ are derived from the conditional variance of the posterior predictive permeability field:
\begin{equation}
    \scriptstyle \mathrm{Var}[K_{NN}^{\,a_h} \,|\, X, \mathcal{D}, \mathcal{M}] 
     \simeq \frac{1}{N_s - N} \sum_{i=N}^{N_s}  
     \left( K_{NN}^{\,a_h}(X;\Theta^{t_i}) - \overline{K_{NN}^{\,a_h}(x,y)} \right)^2
    \label{eq:Var_K}
\end{equation}
which we denote more compactly as $\mathrm{Var}(K_{NN}^{\,a_h}(x,y))$. Darcy flow-based upscaling is performed using the two permeability tensors 
$\mathbf{K} = \overline{K_{NN}^{\,a_h}(x,y)} \pm \sqrt{\mathrm{Var}\!\left(K_{NN}^{\,a_h}(x,y)\right)}$, 
yielding bounds on the upscaled corrected permeability. Table \ref{tableS2} compares these bounds with uncertainty intervals directly estimated from the Bayesian inference and shows good agreement between both approaches across various JRCs. 

Models incorporating corrected local permeability maps derived from hydraulic apertures consistently yield estimates that align more closely with Stokes effective permeability, particularly in rougher fracture profiles. This suggests that including detailed local corrections to the permeability fields enhances the robustness of upscaling approaches across different fracture geometries and roughness patterns, and is therefore highly relevant for larger-scale upscaling, particularly in complex fracture networks.

\section*{Acknowledgements}
This work is funded by the Engineering and Physical Sciences Research Council's ECO-AI Project grant (reference number EP/Y006143/1), with additional financial support from the PETRONAS Center of Excellence in Subsurface Engineering and Energy Transition (PACESET).

\end{document}